\documentclass[12pt,preprint]{aastex}
\usepackage{spr-astr-addons}
\usepackage{lipsum,amsmath,multicol}
\usepackage{amsmath,amssymb}

\RequirePackage{color}

\begin{document}

\title{Stellar objects in the quadratic regime}
\slugcomment{}
\shorttitle{Short article title}
\shortauthors{Autors et al.}

\author{P. Mafa Takisa} 
\affil{Astrophysics and Cosmology Research Unit, School of Mathematics, Statistics and Computer Science, University of KwaZulu-Natal, Private Bag X54001, Durban 4000, South Africa
}

\author{S.D. Maharaj}
\affil{Astrophysics and Cosmology Research Unit, School of Mathematics, Statistics and Computer Science, University of KwaZulu-Natal, Private Bag X54001, Durban 4000, South Africa.}

\author{Subharthi Ray}
\affil{Astrophysics and Cosmology Research Unit, School of Mathematics, Statistics and Computer Science, University of KwaZulu-Natal, Private Bag X54001, Durban 4000, South Africa.}


\begin{abstract}
We model a charged anisotropic relativistic star with a quadratic equation of state. Physical features of an exact solution of the Einstein-Maxwell system are studied by incorporating the effect of the nonlinear term from the equation of state. It is possible to regain the masses, radii and central densities for a linear equation of state in our analysis. We generate masses for stellar compact objects and perform a detailed study of PSR J1614-2230 in particular.
We also show the influence of the nonlinear equation of state on physical features of the matter distribution. We demonstrate that it is possible to incorporate the effects of charge, anisotropy and a quadratic term in the equation of state in modelling a compact relativistic body.


\end{abstract}

\section{Introduction}\label{Sec:intro}
 The Einstein-Maxwell field equations play a significant role in several different applications in relativistic astrophysics.
 The study of \cite{bonnor1965} showed that the electric field is essential in describing the equilibrium of compact bodies, and charge may even halt gravitational collapse.
The necessary requirement in relativistic astrophysics is to build stable equilibrium solutions of the
Einstein-Maxwell system, and to generate models of different  astrophysical objects with strong gravitational fields by choosing appropriate matter distributions. Models built in this way maybe useful in describing the physical characteristics of compact stellar objects such as gravastars, neutron stars, quark stars, etc.
\cite{glendening} has observed that all macroscopic bodies are charge neutral or they may possess a small amount of charge, so that  the structure of the stellar body is not much affected in the latter stages of its evolution. Nevertheless, note that there are early phases in the evolution of a compact star, in which immediate charge neutrality is not possible, but it is attained in later stages. This situation arises, for example, at the birth of a compact star from the core collapse supernova. In this case the electromagnetic field substantially affects the structure of the star.  
A large charge distribution can disrupt the structure of the stellar body.
 \cite{ray2003} showed from the balance of forces and the strength of their coupling, that such huge charge disrupting the stellar structure, has very little
 effect in the equation of state of the content. Taking the protons as the carrier of the charge in the stellar body, they demonstrated that one extra proton in a sea of $10^{18}$ baryons, can generate a total charge that will modify the stars structure.
It is important to point out, that there has been considerable recent improvements in observations relating to compact stellar bodies. The mass of many compact stellar bodies have been found with a fair degree of precision; the main challenge is to accurately determine the radius of the star. Some recent studies have developed improved techniques which give the accurate mass and the radius of some compact stellar objects. This improved observational information about such compact objects has generated much interest about the internal matter content and consequently the spacetime geometry. 

Solutions of the Einstein-Maxwell system with an equation of state are desirable in the description of realistic astrophysical objects. The importance of an equation of state in a stellar model has been comprehensively investigated by \cite{Varela2010}.
Studies in quark stars involve physics at high densities described by a linear equation of state. Various aspects of the phenomenological MIT Bag model with linear quark equation of state have been considered by \cite{witten1984}, \cite{chodos1974} and \cite{weber2005}.
 There have been some attempts made recently to find exact analytic solutions of the Einstein-Maxwell system for bound matter configurations with a linear equation of state. These include the treatments of \cite{ivanov2002}, \cite{sharma2007a}, \cite{thirukkanesh2008}, Mafa Takisa and Maharaj (2013) and \cite{thirukkanesh2013}. Models with a quadratic equation of state are rare because of the increased nonlinearity in the field equations. The models of  \cite{feroze2011} and \cite{maharaj2012}  are recent examples that satisfy a quadratic equation of state. 

The study of the structure of compact stars requires understanding of the equation of state describing the stellar matter under extreme conditions. High mass compact stars, reported recently in the literature, provide strong constraints on the properties of ultradense matter beyond the saturation nuclear density. Even for the 1.44 $M_\odot$ mass or so compact stars, we can still debate over the nature of the star, whether it is composed of nuclear matter, or quark matter or a hybrid of the both. For the hybrid scenario, the core is made of quark matter which has a softer equation of state, and the outside is a stiffer nuclear matter equation of state. In nuclear physics or particle physics, it is often quite challenging to find a single equation of state of matter that smoothly matches the quark matter core with the outer nuclear matter. In this context, it is also worth mentioning that \cite{cottam2002} and \cite{zel2006} showed that most of the current equations of state describing quark matter are too soft and so are unable to explain the existence and stability of massive neutron stars. \cite{rodrigues2011} pointed out that only stiff equations of state  describing normal nuclear matter at high densities would be capable of explaining the stability of high compact star masses with $M \sim 2M_\odot$. Consequently a lot of work has been carried out in hybrid stars with modification to the equation of state in the past years in order to address this issue. In our work, we have chosen a quadratic equation of state, which is softer at low densities and stiffer at higher densities, thus accommodating for a hybrid scenario, as we are more interested in finding the exact solution in general relativity.

In this paper we utilize the quadratic equation of state in a class of exact models found by Maharaj and Mafa Takisa (2012) to study physical features. This enables us to consider deviations from the linear case and the changes to observable quantities such as the mass of the star.

We use an exact solution of the Einstein-Maxwell system found earlier by \cite{maharaj2012} to study physical features and show that this model is consistent with observed objects. We intend to study the effects on a compact object arising from nonlinearities in a quadratic equation of state.
In Sect. \ref{Sec:model}, the Einstein-Maxwell field equations are considered and the \cite{maharaj2012} model is presented. Some recent observations are reviewed in Sect. \ref{Sec:recent}.
In Sect. \ref{Sec:analphys}, masses and radii are generated for particular parameter values in the absence of charge. These are presented in Table 1. In Tables 2-4 in Sect. \ref{Sec:analphys1}, we generate masses, radii and central densities for charged and uncharged bodies. The connection to the astrophysical object PSR J1614-2230 is made. Graphical plots of the physical quantities are made in in Sect. \ref{Sec:analphys2}, and we discuss the significance of the quadratic term in the equation of state.

\section{The model}\label{Sec:model}
The line element for a static spherically symmetric interior matter distribution has the form
\begin{equation}
\label{f1} 
ds^{2} = -e^{2\nu} dt^{2} + e^{2\lambda} dr^{2}
+ r^{2}(d\theta^{2} + \sin^{2}{\theta} d\phi^{2}),
\end{equation}
where $\nu=\nu(r)$ and $\lambda=\lambda(r)$ are the potentials. The energy momentum tensor for an anisotropic charged imperfect
fluid sphere is of the form
\begin{eqnarray}
\label{eq:f2} T^{ab}&=&\mbox{diag}\left(-\rho -\frac{1}{2}E^2, p_r-\frac{1}{2}E^2, p_t + \frac{1}{2}E^2, \right. \nonumber\\
&& \left. p_t+ \frac{1}{2}E^2 \right),
\end{eqnarray}  
which describes a distribution with anisotropy and charge. The quantities $\rho$, $p_{r}$, $p_{t}$ and  $E$ are the density, radial pressure, tangential pressure and electric field intensity respectively.
For a physically reasonable star we require that the
matter distribution satisfies a barotropic equation of state
$p_r=p_r(\rho)$; the quadratic form is given by
\begin{equation}
\label{helen}
 p_r = \gamma\rho^{2}+\alpha\rho -\beta ,
\end{equation}
where $\beta$, $\alpha$ and $\gamma$ are constants. When $\gamma=0$ then we regain a linear equation of state. The constants $\gamma$ and $\alpha$ constrain the density via the sound speed causality condition ($\rho\leq\frac{1+\alpha}{2\gamma}$).
The gravitational interactions on the matter distribution and the electromagnetic field are determined by the Einstein-Maxwell system
\begin{eqnarray}
\label{P10a}
G^{ab} &=&  kT^{ab},\\
\label{P10b}
F_{ab;c}+F_{bc;a}+F_{ca;b} &=& 0,\\
\label{P10c}
{F^{ab}}_{;b} &=& 4\pi J^{a},
\end{eqnarray}
where we have set $k=8\pi$ $(G = c= 1)$ in geometrized units.
The system (\ref{P10a})-(\ref{P10c}) is highly nonlinear and
governs the behaviour of the relativistic star in the presence of the charge.

For the line element (\ref{f1}), the Einstein-Maxwell field equations  (\ref{P10a})-(\ref{P10c}) become
\begin{eqnarray}
\label{f3} 
 8\pi\rho + \frac{1}{2}E^{2}&=&\frac{1}{r^{2}} \left[ r(1-e^{-2\lambda}) \right]',\\
 \label{f4} 
8\pi p_r -\frac{1}{2}E^{2}&=&- \frac{1}{r^{2}} \left( 1-e^{-2\lambda} \right) + \frac{2\nu'}{r}e^{-2\lambda} ,\\
\label{f5} 
8\pi p_t + \frac{1}{2}E^{2}&=&e^{-2\lambda}\left( \nu'' + \nu'^{2}+ \frac{\nu'}{r}\lambda'\right.\nonumber\\ 
&&\left. -\frac{\lambda'}{r} -\nu\right),\\
\label{f6} 
\sigma & = & \frac{1}{4\pi r^{2}} e^{-\lambda}(r^{2}E)', \label{x}
\end{eqnarray}
where $\sigma=\sigma(r)$ is named the proper charge density and primes indicate differentiation with respect to $r$. In the presence of charge the gravitational mass is defined by
\begin{equation}
\label{mass}
M(r)=4\pi\int^{r}_{0} \left(\rho(\omega)_{unch}+\frac{E^{2}}{8\pi}\right)\omega^2 d\omega,
\end{equation}
where  $\rho(\omega)_{unch}$ is the uncharged energy density ($E = 0$).
We remark that equations (\ref{f3})-(\ref{f6}) imply the generalised Tolman-Oppenheimer equation
\begin{equation}
\label{pedr}
 \frac{dp_{r}}{dr}=\frac{2}{r}(p_{t}-p_{r})-r(\rho+p_{r})\nu^{\prime}+\frac{E}{4\pi r^{2}}\left( r^{2}E\right)^{\prime},
\end{equation}
showing that the gradient $\frac{dp_{r}}{dr}$ is influenced by the anisotropy and charge.
The above equation is known as the Bianchi identity representing hydrostatic equilibrium of the charged anisotropic matter. These quantities can drastically change physical quantities such as the surface tension
as shown by \cite{sharma2007b} and \cite{Horvat2009}.

In this paper, we extend the linear treatment of \cite{takisa2014} by investigating the Maharaj and Mafa Takisa (2012) model, and we include the effect of the quadratic term on the structure of observed objects.
On using the quadratic equation of state (\ref{helen}), an exact solution to the Einstein-Maxwell system (\ref{f3})-(\ref{x}) has the form
\begin{eqnarray}
\label{S6}
e^{2\lambda} &=& \frac{1+ar^2}{1+br^2},\\
\label{S7}
e^{2\nu} &=& A^{2}D^{2}\left(1+ar^2 \right)^{2m}[1+br^2]^{2n}\nonumber\\
&&\times\exp[2F(r)],\\
\label{S8}
\rho &=& \frac{(2a-2b)(3+ar^2)-sa^{2}r^{4}}{16\pi(1+ar^2)^{2}},\\
\label{S9}
p_{r} &=& \gamma \rho^{2}+\alpha \rho-\beta ,\\
\label{S10}
p_{t} &=& p_{r}+\Delta,\\
\label{S11}
8\pi\Delta &=&  \frac{4r^2(1+br^2)}{1+ar^2}\left[\frac{m(m-1)a^{2}}{(1+ar^2)^{2}}\right.\nonumber\\    
 &&\left.+\frac{2mnab}{(1+ar^2)(1+br^2)} + \frac{ma{F}'(r)}{r(1+ar^2)}\right.\nonumber\\
& & +\left.\frac{b^{2}n(n-1)}{(1+br^2)^{2}} + \frac{nbF'(r)}{r(1+br^2)}\right.\nonumber\\
&& \left. + \frac{F''(r)}{2r^2}-\frac{F' (r)}{2r^3}+ \frac{F'(r)^{2}}{4r^2}\right]\nonumber\\
&&+\left[-\frac{2(a-b)r^2}{(1+ar^2)^{2}}+\frac{4(1+br^2)}{(1+ar^2)}\right]\nonumber\\
&&\times \left[\frac{am}{1+ar^2}+ \frac{bn}{1+br^2} + \frac{F'(r)}{2r} \right]\nonumber\\
&&-\frac{\gamma}{32 \pi}\left[\frac{(a-b)(3+ar^2)-sa^{2}r^{4}}{2(1+ar^2)^{2}}\right]^{2}\nonumber\\
& & -\frac{1}{2(1+ar^2)^{2}}[2(a-b)+sa^{2}r^{4}\nonumber\\
&&-16 \pi\beta(1+ar^2)^{2}]\nonumber\\
& & -\alpha\left[\frac{(2(a-b)(3+ar^2)-sa^{2}r^{4}}{2(1+ar^2)^{2}}\right],\\
\label{S12}
E^{2}&=& \frac{sa^{2}r^{4}}{(1+ar^2)^{2}},\\
\label{S13}
\sigma^{2} &=& \frac{4sa^2 x (1+br^2)(2+ar^2)^{2}}{\pi (1+ar^2)^{5}}.
\end{eqnarray}
The mass function is
\begin{eqnarray}
\label{S14}
M(r) &=&  \frac{1}{8}\left[\frac{(4(a-b)r^3}{(1+ar^2)}\right.\nonumber\\
&& \left. +\frac{sr(-15-10ar^2+2a^{2}r^{4})}{3a(1+ar^2)}\right.\nonumber\\
&& \left. +\frac{5s\arctan(\sqrt{ar^2})}{a^{3/2}}\right].
\end{eqnarray}
In the above $A$, $a$, $b$, and $s$ are constants. The quantities \textit{F(r)} and the constants \textit{m} and \textit{n} are given by
\begin{eqnarray}
F(r) &=&  \gamma \left[\frac{2(2b-a)(1+ar^2)+(b-a)}{2(1+ar^2)^{2}}\right]\nonumber\\
&& -s\gamma\left[ \frac{(a-b)^{2}(ar^2+2)}{4(a-b)(1+ar^2)^{2}}\right.\nonumber\\
&& \left. -\frac{a(2a+s)(1+ar^2)}{4(a-b)(1+ar^2)^{2}}\right]\nonumber\\
&& -s\gamma\left[\frac{s(a-b)+3sb(1+ar^2)}{32(a-b)^{2}(1+ar^2)^{2}} \right]\nonumber\\
& & +\frac{ar^2}{16b}[s^{2}\gamma -2s(1+\alpha)-4\beta],\label{SF}\nonumber\\
m &=& -\frac{s(1+\alpha)}{8(b-a)}+\frac{\alpha}{2}\nonumber\\
&&+\gamma  [2(a-b)]^{2}
\left[\frac{b^{2}}{(b-a)^{3}}+\frac{b}{(b-a)^{2}}+ \frac{1}{4}\right]\nonumber\\
& & +\frac{s\gamma}{8(a-b)^{3}}\left[(a-b)[2s(a-b)+a+b]\right.\nonumber\\
&&\left. -6ab^{2}+2b^{3}(2a-1) \right],\nonumber\\
n &=& \frac{(1+\alpha)(a-b)}{4b}-\frac{2\alpha(a-b)}{4(b-a)}
+ \frac{\beta (a-b)}{4b^{2}}\nonumber\\
&& +\gamma  [2(a-b)]^{2}\left[\frac{b^{2}}{(b-a)^{3}}+\frac{b}{(b-a)^{2}}+
\frac{1}{4}\right]\nonumber\\
&& +\frac{s\gamma}{16b^{2}(b-a)^{3}}\left[a^{4}(s+4b)\right.\nonumber\\
&&\left. +2b(6a^{2}b^{2}-2a^{3}b)\right] +\frac{sa^{2}(1+\alpha)}{8b^{2}(b-a)}.\nonumber
\end{eqnarray}
The exact solution (\ref{S6})-(\ref{S13}) of the Einstein-Maxwell system is expressed in terms of
elementary functions, which helps in the physical analysis.

We observe that the parameters $a$, $b$, $s$, have the dimension of $length^{-2}$.
This suggests that in numerical calculations we should utilise the following transformations:
\[\tilde{a}=a {\cal L}^{2},~~\tilde{b}=b{\cal L}^{2},~~\tilde{s}=s {\cal L}^{2},\] where ${\cal L}$ is a parameter with dimension of $length$.
The requirements for a physically relevant star, in the absence of charge, are given by \cite{delgaty1998}; the conditions for a charged star were considered by \cite{Fatema2013} and \cite{Murad2013}.

The values of $\tilde{a}$, $\tilde{b}$, $\tilde{s}$ should be chosen in the way that the charged, anisotropic star is well behaved. The energy density $\rho$ should be positive inside the stellar object. The radial pressure $p_{r}$ should vanish at the boundary of the sphere where $p_{r}(\varepsilon)=0$ and $\varepsilon$ is the boundary. The tangential pressure $p_{t}$ should be positive in the interior of the stellar object. The gradient of pressure $\frac{dp_{r}}{dr}<0$ in the interior of the stellar object. The speed of sound should respect the condition $v^{2}=\frac{dp_{r}}{d\rho}\leq1$. At the centre the radial pressure and the tangential pressure should be equal ($p_{r}(0)=p_{t}(0)$) and the measure of anisotropy should vanish ($\Delta(0)=0$). The metric functions $e^{2\lambda}$, $e^{2\nu}$ and the electric field intensity $E$ should  remain positive and regular in the interior of the stellar object.
At the centre the $\rho(0)=\rho_{c}$ must be finite and positive.
The energy condition $\rho-p_{r}-2p_{t}>0$ should be satisfied within the interior of the stellar object. At the boundary $r=\varepsilon$ we require
\begin{eqnarray}
 e^{2\nu(\varepsilon)}&=&1-\frac{2{\cal M}}{\varepsilon}+\frac{Q^{2}}{\varepsilon^{2}},\nonumber\\
\label{S105}
e^{2\lambda(\varepsilon)}&=&\left(1-\frac{2{\cal M}}{\varepsilon}+\frac{Q^{2}}{\varepsilon^{2}}\right)^{-1},\nonumber\\
M(\varepsilon)&=&{\cal M}, \nonumber
\label{S106}
\end{eqnarray}
for continuity of the potentials.

The exterior spacetime is described by the Reissner-Nordstr\"om metric
\begin{eqnarray}
\label{Reissner} 
ds^{2} &=& -\left(1-\frac{2\cal M}{r}+\frac{Q}{r^2}\right) dt^{2} \nonumber\\
&&+ \left(1-\frac{2\cal M}{r}+\frac{Q}{r^2}\right)^{-1} dr^{2}\nonumber\\
&& + r^{2}(d\theta^{2} + \sin^{2}{\theta} d\phi^{2}).
\end{eqnarray}
The spacetimes (\ref{f1}) and (\ref{Reissner}) must match smoothly at the stellar boundary $r=\varepsilon$. Then from the above we obtain the conditions

\begin{eqnarray}
\label{Reissner1}
 1-\frac{2{\cal M}}{\varepsilon}+\frac{Q^{2}}{\varepsilon^{2}}&=& \frac{1+a\varepsilon^2}{1+b\varepsilon^2},\\
\label{Reissner2}
\left(1-\frac{2{\cal M}}{\varepsilon}+\frac{Q^{2}}{\varepsilon^{2}}\right)^{-1}
&=& A^{2}D^{2}\left(1+ar^2 \right)^{2m}\nonumber\\
&&\times[1+br^2]^{2n} \nonumber\\
&&\times\exp[2F(r)].
\end{eqnarray}
This is a system of two equations in the parameters $a$, $b$, $s$, $A$, $D$, $m$, $n$, $\alpha$, $\beta$ and $\gamma$ for a specified radial distance $\varepsilon$. There is sufficient freedom in the parameters to ensure that (\ref{Reissner1})-(\ref{Reissner2}) is always satisfied.

\section{Recent observations}\label{Sec:recent}
\cite{jacoby2005} and  \cite{verbiest2008} utilised the detection of the general relativistic Shapiro
delay to calculate the masses of both the neutron star and its companion in a binary system. This was done to a high degree of precision.  \cite{demorest2010} followed this approach to make  radio timing observations of the binary
millisecond pulsar PSR J1614-2230, which displayed a strong Shapiro delay signature. They implied that the pulsar mass $1.97 \pm 0.08M_\odot$ is the highest mass measured to date with accurate precision. \cite{freire2011} utilised the Arecibo and Green Bank radio timing observations to make a very precise measurement of the apsidal motion, and found new constraints on the orbital orientation of the binary system. This was done in the content of a full determination of the relativistic Shapiro delay. With the help of a comprehensive analysis, they found new restrictions on the mass of the pulsar PSR J1903+0327 and its companion and presented it's accurate mass as $1.667 \pm 0.02M_\odot$.
Recently  \cite{rawls2011} presented an improved method for determining the mass of neutron stars in eclipsing X-ray pulsar binaries. They used a numerical code based on Roche geometry which they supplemented with new spectroscopic and photometric data for 4U 1538-52. This allowed for more accurate modelling of the eclipse duration leading to improved values for the neutron star masses: $1.77 \pm 0.08M_\odot$ for Vela X-1, $1.29 \pm 0.05M_\odot$ for LMC X-4  and $1.29 \pm 0.08M_\odot$ for Cen X-3.

There have been similar observations for other stars.
However for this investigation, we restrict ourselves to pulsar PSR J1614-2230, a binary millisecond large pulsar.

\section{Stellar masses}\label{Sec:analphys}
In this section, we use the analytical solutions (\ref{S6})-(\ref{S14}), with the quadratic equation of state (\ref{S9}), to study the effect of the quadratic term $\gamma$ on the model. We wish to compare the outputs to the recent results of \cite{takisa2014} who considered the linear case. We choose $\gamma$ such that the causality condition $v^{2}=\frac{dp_{r}}{d\rho}\leq1$ is satisfied and take the parameter values: $\tilde{a}=53.34$, ${\cal L}=43.245~ {\rm km}$, $\alpha=0.33$, $\beta=0.5\alpha\times 10^{15}{\rm g}~{\rm cm}^{-3}$ and $\tilde{s}=0.0$.
We are concerned here with uncharged bodies. The parameter 
$\alpha$ has the fixed value $\alpha=\frac{1}{3}$ but the quadratic parameter $\gamma$ is allowed to vary.
We obtain different masses, radii and central densities for different parameter values. The results are given in Table~1. We note that the compactification factor is in the range of $\frac{M}{R}$ $\sim\frac{1}{10}$ to $\frac{1}{4}$; this corresponds to neutron stars and ultra-compact stars as pointed out by \cite{takisa2013}. We find that a variety of stellar masses are generated which correspond to acceptable values of the central density $\rho_{c}$ and the $\frac{M}{R}$ ratio. Of particular interest are the values $\gamma=0.140$, $\alpha=0.33$, $R=10.30$, $\frac{M}{R}=0.191$ and $\rho_{c}=3.45 \times 10^{15}{\rm g}~{\rm cm}^{-3}$ which give the corresponding mass of the PSR J1614-2230. These values are underlined in Table~1. Therefore this astronomical object is consistent with a quadratic equation of state. Note that the same mass is contained in the analysis of \cite{takisa2014} with a linear equation of state.

\begin{table*}
\caption{Variation of mass, radius and central density in term of $\gamma$ in the absence of charge. The parameter $\gamma$ is variable and $\alpha$ is fixed. \label{table1-regular}}
\begin{center}
\begin{tabular}{@{}ccccccccc@{}}
\tableline
$\gamma$&$\tilde{a}-\tilde{b}$&$\tilde{s}$ &$\alpha$&$M$ &$M/R$& $R$(km) &  $\rho_{c}(\times 10^{15}~\mbox{gcm}^{-3})$\\
\tableline
0.100&46.44&0.0 & 0.33& 2.55 & 0.230&11.07 &4.0\\
0.126&44.60&0.0 & 0.33 & 2.37&0.218&10.85 &3.84\\
0.132&42.50&0.0 & 0.33&2.18 &0.206 &10.60 & 3.66\\
\underline{0.140}& \underline{40.01}&0.0&\underline{0.33} &\underline{1.97} & \underline{0.191} &\underline{10.30} &\underline{3.45}\\
0.148&37.73 &0.0& 0.33 &1.77 &0.177&9.99 &3.25\\
0.154&36.47 &0.0& 0.33 &1.667 & 0.170 &9.82 &3.14\\
0.163&34.30 &0.0& 0.33 &1.49 & 0.157&9.51 &2.95\\
0.177& 31.62&0.0& 0.33 &1.29 &0.141& 9.13 &2.72\\
0.189& 29.70&0.0& 0.33 &1.14 & 0.129& 8.83 &2.55\\
0.196& 28.61 &0.0& 0.33 &1.07 & 0.124&8.65 &2.46\\
0.200& 24.92 &0.0& 0.33 &0.89 & 0.111& 8.04 &2.14 \\
\tableline
\end{tabular}
\end{center}
\end{table*}
\section{The pulsar J1614-2230}\label{Sec:analphys1}
The analysis of \cite{takisa2014} was shown to be consistent with observational objects such as Vela X-1, SMC X-1, Cen X-3, PSR J1903+327 and PSR J1614-2230. Our intention in this treatment is to focus on the particular object PSR J1614-2230 $(1.97 \pm 0.08M_\odot)$, since this mass is so far the highest yet measured with accurate precision. We have shown in Sect. \ref{Sec:analphys} that PSR J1614-2230 is consistent with a nonlinear equation of state. Our analysis can be similarly applied to other pulsar objects.
We compute the quantities $M$, $\frac{M}{R}$, $R$ and $\rho_{c}$ by allowing the parameters $\gamma$ and $\alpha$ to be variable. The relevant values are contained in Tables 2-4. 
The masses in Table 2 are uncharged whilst the masses in Table 3 and Table 4 are charged. The underlined values in these tables represent the corresponding values that we expect for the object PSR J1614-2230 when $\gamma=0$. 

Tables 2-4 have been generated with the objective of regaining a central density of $\rho_{c}=3.45$ which is associated with PSR J1614-2230. In Table 2 the quantities $M$, $\frac{M}{R}$ and $R$ have been found in the absence of charge. Different values of the parameters $\gamma$ and $\alpha$ produce uncharged massive objects which are physically reasonable. In Table 3 we similarly generate stellar structures which are reasonable in the presence of charge $\tilde{s}=7.5$. The values in Table 4 are also consistent with observations but with a higher value for the charge $\tilde{s}=14.5$. The presence of charge in Table 3 and Table 4 also produces physically reasonable charged objects for different values of the parameters $\gamma$ and $\alpha$. 
The presence of charge has an effect on the mass and the radius of the object.

\begin{table*}
\caption{Different masses  and radii for PSR J1614-2230 for the uncharged case. The parameters $\gamma$ and $\alpha$ are variable. \label{table2-regular}}
\begin{center}
\begin{tabular}{@{}ccccccccc@{}}
\tableline
$\gamma$ &$\tilde{a}-\tilde{b}$&$\tilde{s}$ &$\alpha$ &$M$ &$M/R$&$R$(km) & $\rho_{c}(\times 10^{15}~\mbox{gcm}^{-3})$\\
\tableline
\underline{0.0}&\underline{40.01}&0.0 & \underline{0.99}&\underline{1.97}&\underline{0.191}&\underline{10.30}& \underline{3.45} \\
0.140 &40.01 &0.0& 0.33&1.97&0.191  & 10.30& 3.45 \\
0.158& 40.01&0.0& 0.24&2.02 & 0.192 &10.50 & 3.45 \\
0.163&40.01&0.0&  0.21  &2.06& 0.192  &10.70 & 3.45 \\
0.177& 40.01&0.0& 0.15 &2.10 &0.193 &10.90 &  3.45 \\
0.196& 40.01& 0.0&0.06 &2.13 &0.193  &11.06&  3.45 \\
0.200& 40.01& 0.0&0.04 &2.14 &0.193 &11.09&  3.45\\
\tableline
\end{tabular}
\end{center}
\end{table*}

\begin{table*}
\caption{Different masses of PSR J1614-2230 and radius for the charged case. The parameters $\gamma$ and $\alpha$ are variable.\label{table3-regular}}
\begin{center}
\begin{tabular}{@{}ccccccccc@{}}
\tableline
$\gamma$&$\tilde{a}-\tilde{b}$&$\tilde{s}$ &$\alpha$&$M$ &$M/R$&$R$(km) & $\rho_{c}(\times 10^{15}~\mbox{gcm}^{-3})$\\
\tableline
\underline{0.0}& \underline{40.01}&7.5& \underline{0.99}&\underline{1.98}&\underline{0.204}&\underline{9.67}& \underline{3.45} \\
0.140&40.01 &7.5& 0.33&1.98&0.204 &9.67& 3.45 \\
0.158&40.01&7.5& 0.24&2.07 &0.205 &10.07& 3.45   \\
0.163& 40.01&7.5& 0.21 &2.13& 0.205  &10.37 & 3.45  \\
0.177&40.01& 7.5&0.15 &2.18 &0.206 &10.56 &  3.45 \\
0.196&40.01& 7.5& 0.06 &2.19 &0.206  &10.65&  3.45 \\
0.200&40.01& 7.5& 0.04&2.22 &0.207 &10.74&  3.45 \\
\tableline
\end{tabular}
\end{center}
\end{table*}

\begin{table*}
\caption{Different masses and radii for PSR J1614-2230 for the charged case. The parameters $\gamma$ and $\alpha$ are variable. \label{table4-regular}}
\begin{center}
\begin{tabular}{@{}ccccccccc@{}}
\tableline
$\gamma$&$\tilde{a}-\tilde{b}$&$\tilde{s}$ &$\alpha$ &$M$ &$M/R$&$R$(km) & $\rho_{c}(\times 10^{15}~\mbox{gcm}^{-3})$\\
\tableline
\underline{0.0}&\underline{40.01}& 14.5&\underline{0.99}&\underline{2.13}&\underline{0.231}&\underline{9.21}& \underline{3.45} \\
0.140 & 40.01&14.5&0.33&2.13&0.231  & 9.21& 3.45  \\
0.158&40.01& 14.5&0.24&2.32 & 0.232 &10.05 & 3.45   \\
0.163&40.01&14.5&0.21  &2.34& 0.232  &10.10 & 3.45  \\
0.177&40.01&14.5&0.15  &2.35 &0.232 &10.15 &  3.45 \\
0.196&40.01&14.5& 0.06 &2.36 &0.232  &10.18&  3.45  \\
0.200&40.01&14.5& 0.04&2.36 &0.232 &10.19&  3.45 \\
\tableline
\end{tabular}
\end{center}
\end{table*}

\section{Discussion}\label{Sec:analphys2}
This physical analysis is completely new, extending the \cite{takisa2014} result and showing interesting features that arise when the quadratic term is present in the equation of state.
To illustrate the effect of the quadratic term of the equation of state with $\gamma\neq0$ in the interior of PSR J1614-2230, we have plotted the energy density $\rho$, radial pressure $p_{r}$, tangential pressure  $p_{t}$, the measure of anisotropy $\Delta$, speed of sound  $v^{2}=\frac{dp_{r}}{d\rho}$, and the quantity $\rho-p_{r}-2p_{t}$ in Figures 1-6 respectively. The presence of $\gamma$ has only a slight effect on the radial pressure, the tangential pressure and the measure of anisotropy profiles. The profiles in the presence of charge and non-zero $\gamma$ are similar to the \cite{takisa2014} analysis with a linear equation of state. The speed of sound is positive and decreasing throughout the star and the causality condition is maintained in Figure 5. The variation $\rho-p_{r}-2p_{t}$ is positive within the star in Figure 6 and the energy condition is satisfied. For $E=0$ and $\gamma=0.200
$, we point out a small of increase for the radius and the mass of  4$\%$ and 4.5$\%$ respectively.
The corresponding results for the charged ($E\neq0$) case are given in Table 3 and Table 4. For the maximum value of $\gamma=0.200$, we note the increase of 8$\%$ for the radius and 9$\%$ for the mass with $E=7.5$. It is clear that the quadratic term $\gamma$ leads to an increase of 11$\%$ in the mass of a stellar object for the maximum value of $E=14.5$. We observe that for both cases $E=0$ and $E\neq0$, the quadratic term $\gamma$ has the effect of increasing the compactification factor $\frac{M}{R}$ slightly. We have shown the relevance of the quadratic equation of state to relativistic objects, in particular to the observed object PSR J1614-2230.

\begin{figure}[h!]
 \centering
 \includegraphics[width=8cm,height=5cm]{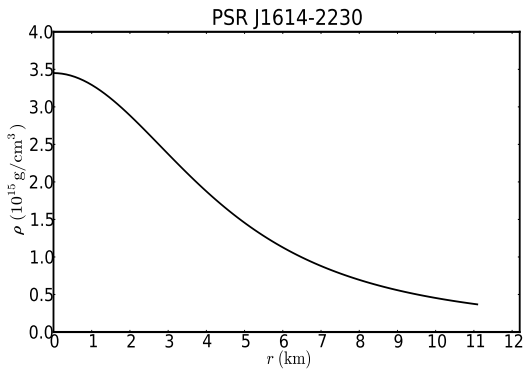}
  \caption{Energy density $\rho(r)$ versus radius.}
 \label{one}
\end{figure}

\begin{figure}[h!]
 \centering
 \includegraphics[width=8cm,height=5cm]{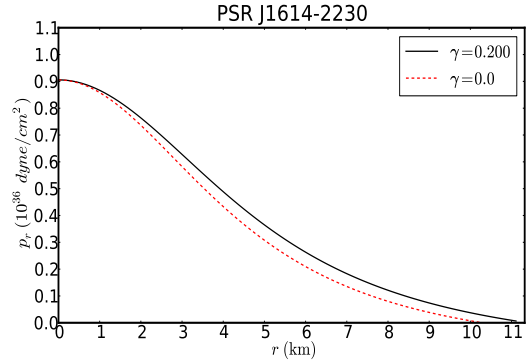}
  \caption{Radial pressure $p_{r}$ versus radius.}
 \label{one}
\end{figure}

\begin{figure}[h!]
 \centering
 \includegraphics[width=8cm,height=5cm]{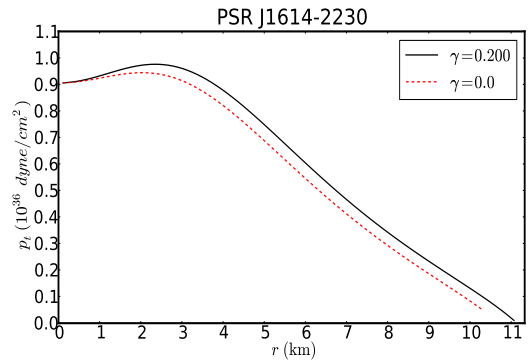}
  \caption{Tangential pressure $p_{t}$ versus radius.}
 \label{one}
\end{figure}

\begin{figure}[h!]
 \centering
 \includegraphics[width=8cm,height=5cm]{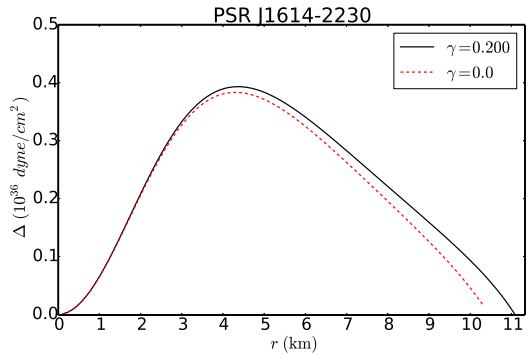}
  \caption{Anisotropy $\Delta$ versus radius.}
 \label{one}
\end{figure}

\begin{figure}[h!]
 \centering
 \includegraphics[width=8cm,height=5cm]{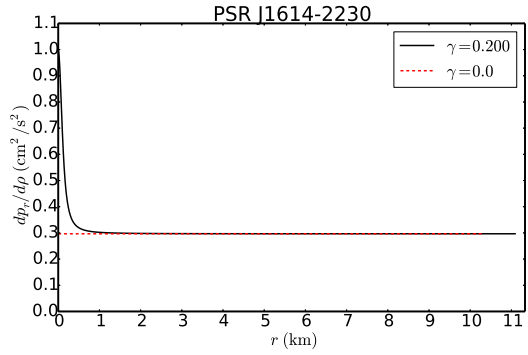}
  \caption{Speed of sound $dp_{r}/d\rho$ versus radius.}
 \label{one}
\end{figure}

\begin{figure}[h!]
 \centering
 \includegraphics[width=8cm,height=5cm]{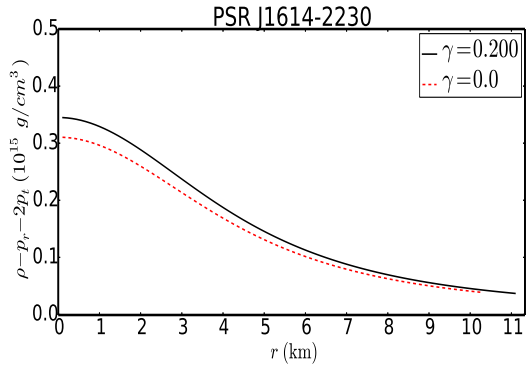}
  \caption{Variation of  $\rho-p_{r}-2p_{t}$ versus radius.}
 \label{one}
\end{figure}

\newpage
\begin{center}
\textbf{Acknowledgements }
\end{center}
PMT thanks the National Research Foundation and the University of
KwaZulu-Natal for financial support.
SDM acknowledges that this work is based upon research supported by the South African Research
Chair Initiative of the Department of Science and
Technology and the National Research Foundation. 
SR acknowledges the NRF incentive funding for research support.

\end{document}